\documentclass[prb,aps,twocolumn,amsmath,amssymb,floatfix,showpacs,
superscriptaddress]{revtex4}

\usepackage[dvips]{graphics}
\usepackage{color}
\definecolor{dred}{rgb}{0,0,0.6}

\begin{document}

\title{\textcolor{dred}{Magnetic-non-magnetic superlattice chain with 
external electric field: Spin transport and the selective switching 
effect}}

\author{Moumita Dey}

\affiliation{Theoretical Condensed Matter Physics Division, Saha
Institute of Nuclear Physics, Sector-I, Block-AF, Bidhannagar,
Kolkata-700 064, India}

\author{Santanu K. Maiti}

\email{santanu.maiti@isical.ac.in}

\affiliation{Physics and Applied Mathematics Unit, Indian Statistical
Institute, 203 Barrackpore Trunk Road, Kolkata-700 108, India}

\author{S. N. Karmakar}

\affiliation{Theoretical Condensed Matter Physics Division, Saha 
Institute of Nuclear Physics, Sector-I, Block-AF, Bidhannagar, 
Kolkata-700 064, India}

\begin{abstract}

Based on Green's function formalism, the existence of multiple mobility 
edges in a one-dimensional magnetic-non-magnetic superlattice geometry in 
presence of external electric field is predicted, and, it leads to the 
possibility of getting a metal-insulator transition at multiple values of 
Fermi energy. The role of electric field on electron localization is 
discussed for different arrangements of magnetic and non-magnetic atomic 
sites in the chain. We also analyze that the model quantum system can be 
used as a perfect spin filter for a wide range of energy.

\end{abstract}

\pacs{73.63.Nm, 72.20.Ee, 73.21.-b, 73.63.Rt}

\maketitle

\section{Introduction}

Quantum transport in low-dimensional systems has been a topic of interest 
within the past few decades due to its potential applicability in the field
of nanoscience and nanotechnology. Exploitation of the spin degree of 
freedom adds a possibility of integrating memory and logic into a single
device, leading to remarkable development in the fields on magnetic data 
storage application, device processing technique, quantum 
computation~\cite{wolf}, etc. Naturally a lot of attention has been 
paid to study spin transport in low-dimensional systems both from 
experimental~\cite{expt1,expt2,expt3} and theoretical~\cite{theo1,theo2,
theo3,theo4,shokri2,shokri4,shokri5,shokri6,san3,san4,san5,sannew1,bellucci1,
bellucci2} points of view. 

The understanding of electronic localization in low-dimensional model 
quantum systems is always an interesting issue. Whereas, it is a well 
established fact that in an infinite one-dimensional ($1$D) system with
random site potentials all energy eigenstates are exponentially localized 
irrespective of the strength of randomness due to Anderson 
localization~\cite{anderson}, there exists another kind of localization, 
known as Wannier-Stark localization, which results from a static bias 
applied to a regular $1$D lattice, even in absence of any 
disorder~\cite{wannier}. Till date a large number of works have been
done to explore the understanding of Anderson localization and scaling 
hypothesis in one- and two-dimensional systems~\cite{tvr}. Similarly,
Wannier-Stark localization has also drawn the attention of many 
theorists~\cite{starktheo1,starktheo2,starktheo3,starktheo4,starktheo5} 
as well as experimentalists~\cite{starkexp}. For both these two cases,
viz, infinite $1$D materials with random site energies and $1$D systems
subjected to an external electric field, one never encounters any 
{\em mobility edge} i.e., energy eigenvalues separating localized 
states from the extended ones, since all the eigenstates are localized.
But there exist some special types of $1$D materials, like quasi-periodic
Aubry-Andre model and correlated disordered systems where mobility edge
phenomenon at some particular energies is obtained~\cite{dun,sanch,fa,
fm,dom,aubry,san6,eco,das,rolf}. Although the studies involving mobility 
edge phenomenon in low-dimensional systems have already generated a wealth 
of literature~\cite{eco,das,rolf,sch,san1,san2,sannew} there is still need 
to look deeper into the problem to address several interesting issues those 
have not yet been explored. For example, whether the mobility edges can 
be observed in some other simple $1$D materials or the number of mobility 
edges separating the extended and localized regions in the full energy 
band of an $1$D material can be regulated, are still to be investigated. 

To address these issues in the present article we investigate two-terminal
spin dependent transport in a $1$D mesoscopic chain composed of magnetic
and non-magnetic atomic sites in presence of external electric field. To 
the best of our knowledge, no rigorous effort has been made so far to 
explore the effect of an external electric field on electron transport in
such a $1$D magnetic-non-magnetic superlattice geometry. Here we show that,
depending on the unit cell configuration, a $1$D superlattice structure
subjected to an external electric field exhibits multiple mobility edges 
at different values of the carrier energy. We use a simple tight-binding
(TB) framework to illustrate the model quantum system and numerically
evaluate two-terminal spin dependent transmission probabilities through 
the superlattice geometry based on the Green's function formalism. From our
exact numerical analysis we establish that a sharp crossover from a 
completely opaque to a fully or partly transmitting zone takes place which
leads to a possibility of tuning the electron transport by gating the 
transmission zone. In addition to this behavior we also show that the
magnetic-non-magnetic superlattice structure can be used as a pure spin
filter for a wide range of energy. These phenomena enhance the prospect 
of such simple superlattice structures as switching devices at multiple
energies as well as spin filter devices, the design of which has significant
impact in the present age of nanotechnology.

With an introduction in Section I, we organize the paper as follows.
In Section II, first we present the model, then describe the theoretical
formulation which include the Hamiltonian and the formulation for 
transmission probabilities through the model quantum system. The 
numerical results are illustrated in Section III and finally, in 
Section IV, we draw our conclusions.

\section{Theoretical Framework}

\begin{figure}[ht]
{\centering \resizebox*{8cm}{2.3cm}{\includegraphics{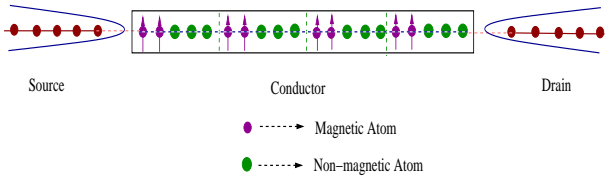}}\par}
\caption{(Color online). A $1$D mesoscopic chain composed of magnetic 
(filled magenta circle) and non-magnetic (filled green circle) atomic 
sites is attached to two semi-infinite $1$D non-magnetic metallic 
electrodes, namely, source and drain.}
\label{chain}
\end{figure}

Let us start with Fig.~\ref{chain} where a $1$D mesoscopic chain composed
of magnetic and non-magnetic atomic sites is attached to two semi-infinite
$1$D non-magnetic electrodes, namely, source and drain. The chain consists 
of $p$ ($p$ being an integer) number of unit cells in which each unit cell 
contains $n$ and $m$ numbers of magnetic and non-magnetic atoms, respectively.
Both the chain and side-attached electrodes are described by simple 
TB framework within nearest-neighbor hopping approximation. 

The Hamiltonian for the entire system can be written as a sum of three 
terms as,
\begin{equation}
H= H_{c} + H_{l} + H_{tun}.
\label{eqn1}
\end{equation}
The first term represents the Hamiltonian for the chain and it reads
\begin{equation}
H_{c} = \sum_i \mbox{\boldmath $c$}_{i}^{\dag} 
(\mbox{\boldmath $\epsilon$}_i + 
\mbox{\boldmath $\vec{h}_i.\vec{\sigma}$}) 
\mbox{\boldmath $c$}_i + \sum_{i} \left[\mbox{\boldmath $c$}_{i}^{\dag} 
\mbox{\boldmath $t$} \mbox{\boldmath $c$}_{i+1} + h.c.\right] 
\label{eqn2}
\end{equation}
where,
$\mbox{\boldmath $c$}^{\dagger}_{i}=\left(\begin{array}{cc}
c_{i\uparrow}^{\dagger} & c_{i\downarrow}^{\dagger} 
\end{array}\right);$
$\mbox{\boldmath $c$}_{i}=\left(\begin{array}{c}
c_{i\uparrow} \\
c_{i\downarrow}\end{array}\right);$
$\mbox{\boldmath $\epsilon$}_i=\left(\begin{array}{cc}
\epsilon_i & 0 \\
0 & \epsilon_i \end{array}\right)$; \\
$\mbox{\boldmath $t$}=t\left(\begin{array}{cc}
1 & 0 \\
0 & 1 \end{array}\right);$ \mbox{and}\\
{\boldmath $\vec{h_i}.\vec{\sigma}$} = $h_i\left(\begin{array}{cc}
\cos \theta_i & \sin \theta_i e^{-j \phi_i} \\
\sin \theta_i e^{j \phi_i} & -\cos \theta_i \end{array}\right)$. \\
~\\
Here, $\epsilon_i$ refers to the on-site energy of an electron at the
site $i$ with spin $\sigma$ ($\uparrow,\downarrow$), $t$ is the 
nearest-neighbor hopping strength, $c_{i\sigma^{\dagger}}$ ($c_{i\sigma}$) 
is the creation (annihilation) operator of an electron at the $i$th site 
with spin $\sigma$ and $h_i$ is the strength of local magnetic moment where 
$h_i=0$ for non-magnetic sites. The term {\boldmath $\vec{h_i}.\vec{\sigma}$} 
corresponds to the interaction of the spin of the injected electron with the 
local magnetic moment placed at the site $i$. The direction of magnetization 
in each magnetic site is chosen to be arbitrary and specified by angles 
$\theta_i$ and $\phi_i$ in spherical polar co-ordinate system for the $i$th 
atomic site. Here, $\theta_i$ represents the angle between the direction of 
magnetization and the chosen $Z$ axis, and $\phi_i$ represents the azimuthal 
angle made by the projection of the local moment on $X$-$Y$ plane with the 
$X$ axis. In presence of bias voltage $V$ between the source and drain an 
electric field is developed, and therefore, the site energies of the chain
becomes voltage dependent. Mathematically we can express it as 
$\epsilon_i=\epsilon_i^0 + \epsilon_i(V)$, where $\epsilon_i^0$ is the 
voltage independent term. The voltage dependence of $\epsilon_i(V)$ reflects
the bare electric field in the bias junction as well as screening due to
longer range electron-electron interaction. In the absence of such screening
the electric field varies uniformly along the chain and it reads
$\epsilon_i(V)=V/2-iV/(N+1)$, where $N$ corresponds to the total number of
atomic sites in the chain. In our present work, we consider both the linear
\begin{figure}[ht]
{\centering \resizebox*{6cm}{3.5cm}{\includegraphics{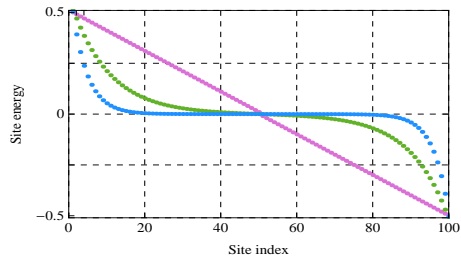}}\par}
\caption{(Color online). Voltage dependent site energies in a $1$D chain
considering $100$ atomic sites for three different electrostatic potential
profiles when the bias voltage $V$ is set equal to $1$.}
\label{potprofile}
\end{figure}
and screened electric field profiles. As illustrative example, in 
Fig.~\ref{potprofile} we show the variation of voltage dependent site 
energies for three different electrostatic potential profiles for a chain 
considering $100$ atomic sites and describe the nature of electronic 
localization for these profiles in the forthcoming section.

The second and third terms of Eq.~\ref{eqn1} describe the TB Hamiltonians 
for the $1$D semi-infinite non-magnetic electrodes and the chain-to-electrode 
coupling. These Hamiltonians are written as follows.
\begin{equation}
H_{l} = \sum \limits_{\alpha = S,D} 
\left[ \sum_n \mbox{\boldmath $c$}_{n}^{\dag} 
\mbox{\boldmath $\epsilon_l$} \mbox{\boldmath $c$}_n 
+ \sum_{n} \left[\mbox{\boldmath $c$}_n^{\dag} 
\mbox{\boldmath $t_l$} \mbox{\boldmath $c$}_{n+1} + h.c.\right] \right]
\label{eqn3}
\end{equation}
and,
\begin{eqnarray}
H_{tun} & = & H_{tun,S} + H_{tun,D} \nonumber \\
& = & \tau_s[\mbox {\boldmath $c$}_1^{\dag} \mbox {\boldmath $c$}_0 
+ h.c.] + \tau_d[\mbox {\boldmath $c$}_N^{\dag} \mbox {\boldmath 
$c$}_{N+1} + h.c.].
\label{eqn4}
\end{eqnarray}
The summation over S and D in Eq.~\ref{eqn3} implies the incorporation of 
both the two electrodes, viz, source and drain. $\epsilon_l$ and $t_l$ 
stand for the site energy and nearest-neighbor coupling, respectively. 
The electrodes are directly coupled to the chain through the lattice sites
$1$ and $N$, and the coupling strengths between these electrodes with
the chain are described by $\tau_s$ and $\tau_d$, respectively.

To obtain spin resolved transmission probabilities of an electron through 
the source-chain-drain bridge system, we use Green's function formalism. 
The single particle Green's function operator representing the entire 
system for an electron with energy $E$ is defined as,
\begin{equation}
G=\left( E - H + i\eta \right)^{-1}
\label{eqn5}
\end{equation}
where, $\eta \rightarrow 0^+$.

Following the matrix form of \mbox{\boldmath $H$} and \mbox{\boldmath $G$} 
the problem of finding \mbox{\boldmath $G$} in the full Hilbert space 
\mbox{\boldmath $H$} can be mapped exactly to a Green's function 
\mbox{\boldmath $G$}$_{c}^{eff}$ corresponding to an effective Hamiltonian 
in the reduced Hilbert space of the chain itself and we have,
\begin{equation}
\mbox{\boldmath ${\mathcal G}$}=\mbox {\boldmath $G$}_{c}^{eff} = \sum 
\limits_{\sigma} \left(\mbox {\boldmath $E$}- 
\mbox {\boldmath $H$}_{c}-\mbox {\boldmath $\Sigma$}_S^{\sigma}-
\mbox {\boldmath $\Sigma$}_D^{\sigma}\right)^{-1},
\label{equ6}
\end{equation}
where, 
\begin{eqnarray}
\mbox{\boldmath $\Sigma$}_{S(D)}^{\sigma} & = & \mbox{\boldmath 
$H$}_{tun,S(D)}^{\dag} \mbox {\boldmath $G$}_{S(D)} 
\mbox{\boldmath $H$}_{tun,S(D)}.
\label{eqn7}
\end{eqnarray}
These \mbox{\boldmath $\Sigma_S$} and \mbox{\boldmath $\Sigma_D$} are the 
self-energies introduced to incorporate the effect of coupling of the chain 
to the source and drain, respectively. Using Dyson equation the analytic 
form of the self energies can be evaluated as follows,
\begin{equation}
\Sigma_{S(D)}^{\sigma} = \frac{\tau_{s(d)}^2}{E - \epsilon_{l}-\xi_{l}}
\label{eqn8}
\end{equation}
where, $\xi_{l} = (E-\epsilon_{l})/2 - i\sqrt{t_{l}^2-(E-\epsilon_{l})^2/4}$.

Following Fisher-Lee relation, the transmission probability of an electron 
from the source to drain is given by the expression,
\begin{equation}
T_{\sigma \sigma^{\prime}} = \mbox{Tr}[\mbox{\boldmath 
$\Gamma$}_{S}^{\sigma} \mbox {\boldmath $\mathcal {G}$}^r 
\mbox {\boldmath $\Gamma$}_{D}^{\sigma^{\prime}} 
\mbox {\boldmath $\mathcal {G}$}^a].
\label{eqn9}
\end{equation}
where, {\boldmath $\Gamma$}$_{S(D)}^{\sigma}$'s are the coupling matrices 
representing the coupling between the chain and the electrodes and they 
are defined as,
\begin{equation}
\mbox {\boldmath $\Gamma$}_{S(D)}^{\sigma} = i \left[\mbox {\boldmath 
$\Sigma$}^{\sigma}_{S(D)} - \mbox {\boldmath 
$\Sigma$}^{\sigma \dag}_{S(D)}\right].
\label{eqn10}
\end{equation}
Here, \mbox{\boldmath $\Sigma$}$_{k}^{\sigma}$ and \mbox{\boldmath 
$\Sigma$}$_{k}^{\sigma \dag}$ are the retarded and advanced self-energies 
associated with the $k$-th ($k=S,D$) electrode, respectively.

Finally, we determine the average density of states (ADOS), $\rho(E)$, 
from the following relation,
\begin{equation}
\rho(E)=-\frac{1}{N \pi} {\mbox{Im}} \left[{\mbox{Tr}} 
[\mbox{\boldmath ${\mathcal G}$}]\right].
\label{equ11}
\end{equation}

In what follows we limit ourselves to absolute zero temperature and use the
units where $c=e=h=1$. For the numerical calculations we set $t=1$, 
$\epsilon_i^0=0 \, \forall \, i$, $h_i=1$ for the magnetic sites, $\theta_i=
\phi_i=0$, $\epsilon_l=0$, $t_l=1$ and $\tau_s=\tau_d=0.8$. The energy 
scale is measured in unit of $t$.

\section {Numerical Results and Discussion}

Throughout our numerical calculations we assume that the magnetic 
moments are aligned along $+Z$ direction ($\theta_i=\phi_i=0$), 
which yields vanishing spin flip transmission probability, viz, 
$T_{\uparrow\downarrow}=T_{\downarrow\uparrow}=0$, across the bridge 
system. The net transmission probability is therefore a sum 
$T(E)=T_{\uparrow\uparrow}(E) + T_{\downarrow\downarrow}(E)$, and the 
origin of this zero spin flipping can be explained from the following
arguments. The operators $\sigma_+$ $(=\sigma_x + i\sigma_y)$ and 
$\sigma_-$ $(=\sigma_x-i\sigma_y)$ associated with the term 
\mbox{\boldmath $\vec{h}_i.\vec{\sigma}$} in the TB Hamiltonian 
Eq.~\ref{eqn2} are responsible for the spin flipping, where 
\mbox{\boldmath $\vec{\sigma}$} being the Pauli spin vector with components 
$\sigma_x$, $\sigma_y$ and $\sigma_z$ for the injecting electron. In our
present model since we consider that all the magnetic moments are aligned 
along $+Z$ direction, the term \mbox{\boldmath $\vec{h}_i.\vec{\sigma}$}
$(=h_{ix}\sigma_x + h_{iy}\sigma_y +h_{iz}\sigma_z)$ gets the form
$h_{iz}\sigma_z$, and accordingly, the Hamiltonian does not contain $\sigma_x$
\begin{figure}[ht]
{\centering \resizebox*{8cm}{11cm}{\includegraphics{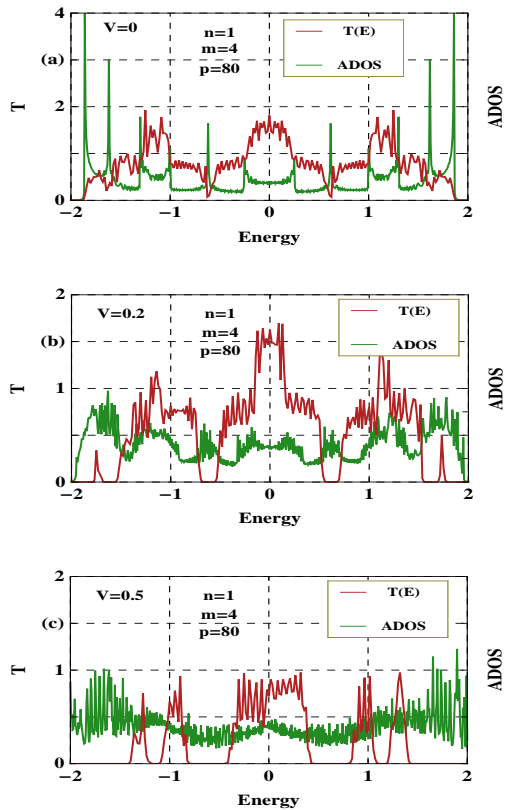}}\par}
\caption{(Color online). Transmission probability $T$ and ADOS as a function 
of energy for a $1$D magnetic-non-magnetic superlattice geometry considering 
a linear bias drop along the chain, as shown by the pink curve in 
Fig.~\ref{potprofile}, where (a)-(c) correspond to the results for three
different values of bias voltage $V$.}
\label{localization1}
\end{figure}
and $\sigma_y$ and so $\sigma_+$ and $\sigma_-$ do not appear, which leads 
to the vanishing spin flip transmission probability across the $1$D chain. 
Below, we address the central results of our study i.e, the possibility of 
getting multiple mobility edges in $1$D magnetic-non-magnetic superlattice 
geometries and how such a simple model quantum system can be used as a 
perfect spin filter for a wide range of energy.

In Fig.~\ref{localization1} we show the variation of total transmission 
probability $T$ along with the average density of states for a $1$D 
magnetic-non-magnetic superlattice geometry considering a linear bias 
drop. Here we consider a $400$-site chain in which each unit cell contains
one magnetic and four non-magnetic sites and the results are shown for three
different bias voltages. For the particular case when the chain is free
from external electric field i.e., $V=0$ electronic conduction through
the bridge takes place for the entire energy band as shown in
Fig.~\ref{localization1}(a) which predicts that all the energy eigenstates 
are extended in nature. The situation becomes really very interesting
when the superlattice geometry is subjected to an external electric field. 
\begin{figure}[ht]
{\centering \resizebox*{8cm}{11cm}{\includegraphics{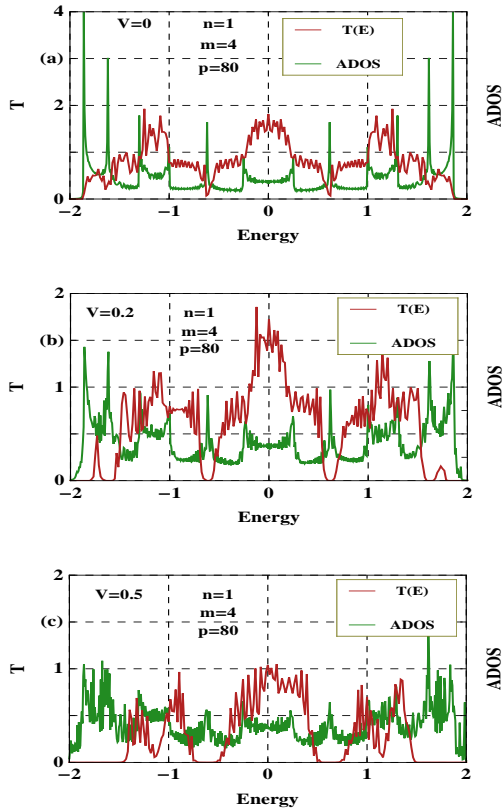}}\par}
\caption{(Color online). Transmission probability $T$ and ADOS as a function
of energy for a $1$D magnetic-non-magnetic superlattice geometry when the
electrostatic potential profile varies following the green curve shown in
Fig.~\ref{potprofile}, where (a)-(c) represent the identical meaning as
in Fig.~\ref{localization1}.}
\label{localization2}
\end{figure}
It is illustrated in Figs.~\ref{localization1}(b) and \ref{localization1}(c). 
From these spectra we notice that there are some energy regions for which 
the transmission probability completely drops to zero which reveals that the 
eigenstates associated with these energies are localized, and they are 
separated from the extended energy eigenstates. Thus, sharp mobility edges 
are obtained in the spectrum, and, the total number of such mobility edges 
separating the extended and localized regions in a superlattice geometry in 
presence electric field strongly depends on the unit cell configuration and 
it can be regulated by adjusting the number of magnetic and non-magnetic 
sites. This phenomenon describes the existence of multiple mobility edges 
in a superlattice geometry under finite bias condition. Now if the Fermi 
energy is fixed at a suitable energy zone where $T$ drops to zero an 
insulating phase will appear, while for the other case, where $T$ is finite, 
a metallic phase is observed and it leads to the possibility of controlling 
the electronic transmission by gating the transmission zone. The width of 
the localized regions between the band of extended regions increases with 
the strength of the electric 
\begin{figure}[ht]
{\centering \resizebox*{8cm}{11cm}{\includegraphics{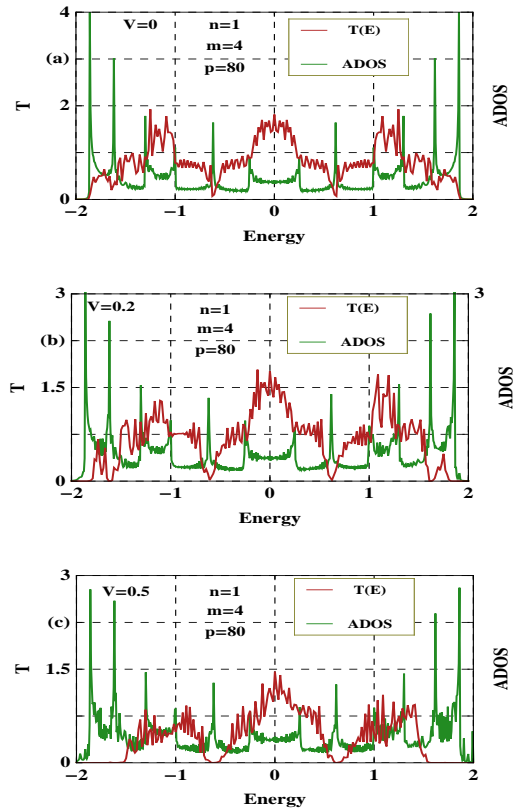}}\par}
\caption{(Color online). Transmission probability $T$ and ADOS as a function
of energy for a $1$D magnetic-non-magnetic superlattice geometry when the
electrostatic potential profile varies following the blue curve shown in
Fig.~\ref{potprofile}, where (a)-(c) represent the identical meaning as
in Fig.~\ref{localization1}.}
\label{localization3}
\end{figure}
field as clearly shown by comparing the spectra given in 
Figs.~\ref{localization1}(b) and \ref{localization1}(c), and, for strong 
enough field strength almost all energy eigenstates are localized. In that 
particular limit metal-to-insulator transition will no longer be observed.

The above results are analyzed for a particular (linear) variation of 
electric field along the chain. To explore the sensitivity of getting 
metal-to-insulator transition on the distribution of electric field, in 
Figs.~\ref{localization2} and \ref{localization3} we present the results 
for two different screened electric field profiles taking the identical 
chain length. From the spectra we clearly observe that the width of the 
localized region gradually disappears with the flatness of the electric 
field profile in the interior of the bridge system. If the potential drop 
takes place only at the chain-to-electrode interfaces, i.e., when the 
potential profile becomes almost flat along the chain the width of the 
localized region almost vanishes and the metal-to-insulator transition is 
not observed, as is the case for the zero bias limit.

Finally, we illustrate how such a simple magnetic-non-magnetic 
superlattice geometry can be utilized as a perfect spin filter for 
\begin{figure}[ht]
{\centering \resizebox*{8cm}{11cm}{\includegraphics{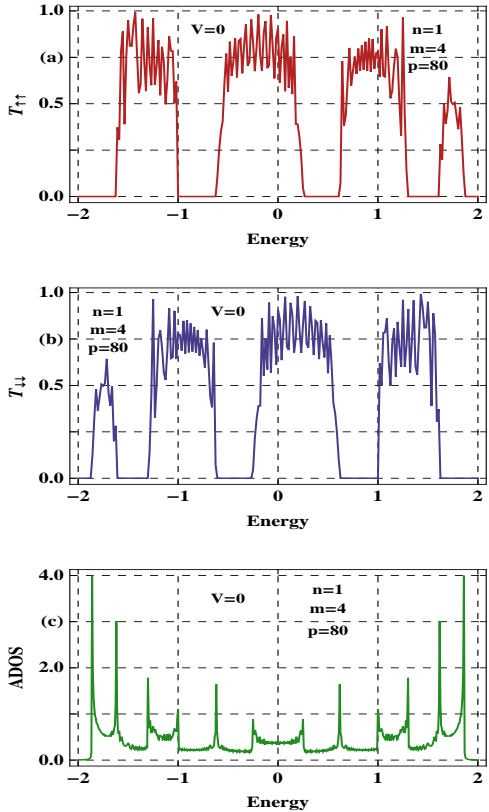}}\par}
\caption{(Color online). $T_{\uparrow\uparrow}$, $T_{\downarrow\downarrow}$
and ADOS as a function of energy for a $1$D magnetic-non-magnetic 
superlattice geometry in absence of external electric field.}
\label{filter1}
\end{figure}
a wide range of energy in absence of any external electric field. As 
illustrative example, in Fig.~\ref{filter1} we present the transmission
probabilities for up and down spin electrons together with the average 
density of states as a function of energy for a $1$D magnetic-non-magnetic 
superlattice geometry. From the spectra we observe that the up and down
spin electrons follow two different channels while traversing through 
the superlattice geometry, since the spin flipping is completely blocked
for this configuration. This splitting of up and down spin conduction 
channels is responsible for spin filtering action and the total number of
these channels strongly depends on the unit cell configuration. From 
Figs.~\ref{filter1}(a) and (b) we clearly see that for a wide range of 
energy for which the transmission probability of up spin electrons drops to 
zero value, shows non-zero transmission probability of down spin electrons.
Therefore, setting the Fermi energy to a suitable energy region we can
control the transmission characteristics of up and down spin electrons,
and, a spin selective transmission is thus obtained through the bridge 
system. Before we end, we would like to point out that since the overlap 
between the up and down spin conduction channels depends on the magnitudes 
of the local magnetic moments, we can regulate the spin degree of 
polarization (DOP) simply by tuning the strength of these magnetic moments 
and for a wide range of energies it (DOP) almost reaches to $100\%$. Thus, 
our proposed magnetic-non-magnetic superlattice geometry is a very good 
example for designing a spin filter.

\section{Conclusion}

To conclude, in the present work we investigate in detail the spin 
dependent transport under finite bias condition through a $1$D
magnetic-non-magnetic superlattice geometry using Green's function
formalism. We use a simple TB framework to describe the model quantum 
system where all the calculations are done numerically. From our exact
numerical analysis we predict that in such a simple $1$D 
magnetic-non-magnetic superlattice geometry multiple mobility edges 
separating the localized and extended regions are obtained in presence 
of external electric field and the total number of mobility edges in 
the full energy spectrum can be controlled by arranging the unit cell 
configuration. This phenomenon reveals that the superlattice geometry 
can be used as a switching device for multiple values of Fermi energy. 
The sensitivity of metal-to-insulator transition and vice versa on the 
electrostatic potential profile is thoroughly discussed. Finally, we 
analyze how such a superlattice geometry can be utilized in designing 
a tailor made spin filter device for wide range of energies. Setting the
Fermi energy at a suitable energy zone, a spin selective transmission is
obtained through the bridge system. All these predicted results may be 
utilized in fabricating spin based nano electronic devices. 

The results presented in this communication are worked out for absolute zero 
temperature. However, they should remain valid even in a certain range 
of finite temperatures ($\sim 300$\,K). This is because the broadening of 
the energy levels of the chain due to the chain-to-electrode coupling is, 
in general, much larger than that of the thermal broadening~\cite{datta1}.


\begin{thebibliography}{99}

\bibitem{wolf} S. A. Wolf, D. D. Awschalom, R. A. Buhrman, J. M. Daughton,
S. von Moln\'{a}r, M. L. Roukes, A. Y. Chtchelkanova, and D. M. Treger, 
Science \textbf{294}, 1488 (2001).

\bibitem{expt1} L. P. Rokhinson, V. Larkina, Y. B. Lyanda-Geller, L. N. 
Pfeiffer, and K. W. West, Phys. Rev. Lett. \textbf{93}, 146601 (2004).

\bibitem{expt2} S. Sahoo, T. Kontos, J. Furer, C. Hoffmann, M. Gr\"{a}ber, 
A. Cottet, and C. Sch\"{o}nenberger, Nature Phys. \textbf{1}, 99 (2005).

\bibitem{expt3} N. Tombros, C. Jozsa, M. Popinciuc, H. T. Jonkman,
and B. J. van Wees, Nature \textbf{448}, 571 (2007).

\bibitem{theo1} I. A. Shelykh, N. T. Bagraev, N. G. Galkin, and
L. E. Klyanchkin, Phys. Rev. B \textbf{71}, 113311 (2005).

\bibitem{theo2} H. W. Wu, J. Zhou, and Q. W. Shi, Appl. Phys. Lett.
\textbf{85}, 1012 (2004).

\bibitem{theo3} D. Frustaglia, M. Hentschel, and K. Richter, Phys.
Rev. Lett. \textbf{87}, 256602 (2001).

\bibitem{theo4} R. Ionicioiu and I. D'Amico, Phys. Rev. B \textbf{67},
041307(R) (2003).

\bibitem{shokri2} A. A. Shokri, M. Mardaani, and K. Esfarjani, Physica E
\textbf{27}, 325 (2005).

\bibitem{shokri4} M. Mardaani and A. A. Shokri, Chem. Phys. \textbf{324}, 
541 (2006).

\bibitem{shokri5} A. A. Shokri and A. Daemi, Eur. Phys. J. B \textbf{69}, 
245 (2009).

\bibitem{shokri6} A. A. Shokri and A. Saffarzadeh, J. Phys.: Condens. 
Matter \textbf{16}, 4455 (2004).

\bibitem{san3} M. Dey, S. K. Maiti, and S. N. Karmakar, Eur. Phys. 
J. B \textbf{80}, 105 (2011).

\bibitem{san4} M. Dey, S. K. Maiti, and S. N. Karmakar, J. Appl. Phys.
\textbf{109}, 024304 (2011).

\bibitem{san5} M. Dey, S. K. Maiti, and S. N. Karmakar, Phys. Lett. A
\textbf{374}, 1522 (2010).

\bibitem{sannew1} M. Dey, S. K. Maiti, and S. N. Karmakar, J. Comput. 
Theor. Nanosci. \textbf{8}, 253 (2011).

\bibitem{bellucci1} S. Bellucci and P. Onorato, Phys. Rev. B \textbf{78},
235312 (2008).

\bibitem{bellucci2} S. Bellucci and P. Onorato, J. Phys.: Condens. Matter 
\textbf{19}, 395020 (2007).

\bibitem{anderson} P. W. Anderson, Phys. Rev. \textbf{109}, 1492 (1958).

\bibitem{wannier} G. H. Wannier, Phys. Rev. \textbf{117}, 432 (1960).

\bibitem{tvr} E. Abrahams, P. W. Anderson, D. C. Licciardello, and 
T. V. Ramakrishnan, Phys. Rev. Lett. \textbf{42}, 673 (1979).

\bibitem{starktheo1} H. M. James, Phys. Rev. \textbf{76}, 1611 (1949).

\bibitem{starktheo2} D. Emin and C. F. Hart, Phys. Rev. B \textbf{36},
7353 (1987).

\bibitem{starktheo3} R. Ouasti, N. Jekri, A. Brezini, and C. Depollier,
J. Phys.: Condens. Matter \textbf{7}, 811 (1995).

\bibitem{starktheo4} N. Zekri, M. Schreiber, R. Ouasti, R. Bouamrane,
and A. Brezini, Z. Phys. B \textbf{99}, 381 (1996).

\bibitem{starktheo5} J. R. Borysowicz, Phys. Lett. A. \textbf{231}, 
240 (1997).

\bibitem{starkexp} C. Hamaguchi, M. Yamaguchi, H. Nagasawa, M. Morifuji, 
A. Di Carlo, P. Vogl, G. B\"{o}hm, G. Tr\"{a}nkle, G. Weimann, 
Y. Nishikawa and S. Muto, Jpn. J. Appl. Phys. \textbf{34}, 4519 (1995).

\bibitem{dun} D. H. Dunlap, H.-L. Wu, and P. Phillips, Phys. Rev. Lett.
\textbf{65}, 88 (1990).

\bibitem{sanch} A. S\'{a}nchez, E. Maci\'{a}, and F. Dom\'{i}nguez-Adame,
Phys. Rev. B \textbf{49}, 147 (1994).

\bibitem{fa} F. A. B. F. de Moura and M. L. Lyra, Phys. Rev. Lett.
\textbf{81}, 3735 (1998).

\bibitem{fm} F. M. Izrailev and A. A. Krokhin, Phys. Rev. Lett.
\textbf{82}, 4062 (1999).

\bibitem{dom} F. Dom\'{i}nguez-Adame, V. A. Malyshev, F. A. B. F. de
Moura, and M. L. Lyra, Phys. Rev. Lett. \textbf{91}, 197402 (2003).

\bibitem{aubry} S. Aubry and G. Andr\'{e}, in {\em Group Theoretical Methods
in Physics}, Annals of the Israel Physical Society Vol. 3, edited by
L. Horwitz and Y. Neeman (American Institute of Physics, New York, 1980),
p. 133.

\bibitem{san6} S. Sil, S. K. Maiti, and A. Chakrabarti, Phys. Rev. B
\textbf{79}, 193309 (2009).

\bibitem{eco} C. M. Soukoulis and E. N. Economou, Phys. Rev. Lett.
\textbf{48}, 1043 (1982).

\bibitem{das} S. Das Sarma, S. He, and X. C. Xie, Phys. Rev. Lett.
\textbf{61}, 2144 (1988).

\bibitem{rolf} M. Johansson and R. Riklund, Phys. Rev. B \textbf{42},
8244 (1990).

\bibitem{sch} A. Eilmes, R. A. R\"{o}mer, and M. Schreiber, Eur. Phys.
J. B \textbf{23}, 229 (2001).

\bibitem{san1} S. Sil, S. K. Maiti, and A. Chakrabarti, Phys. Rev.
Lett. \textbf{101}, 076803 (2008).

\bibitem{san2} S. Sil, S. K. Maiti, and A. Chakrabarti, Phys. Rev. B
\textbf{78}, 113103 (2008).

\bibitem{sannew} S. K. Maiti and A. Nitzan, Phys. Lett. A \textbf{377},
1205 (2013).

\bibitem{datta1} S. Datta, {\em Electronic transport in mesoscopic systems},
Cambridge University Press, Cambridge (1995).

\end{thebibliography}
\end{document}